\documentclass[11pt]{revtex4}

\usepackage{hyperref}
\usepackage{amsmath,amsfonts,amssymb}
\usepackage{braket}
\hypersetup{dvips,dvipdfm,colorlinks=true,urlcolor=magenta,filecolor=magenta,linktoc=page,citecolor=red,linkcolor=blue,bookmarks=true}
\usepackage{graphicx,epstopdf}

\begin{document}
	
	\title{ Classical and quantum cosmology in $f(T)$-gravity theory: A Noether symmetry approach}
	\author{Roshni Bhaumik$^1$\footnote {roshnibhaumik1995@gmail.com}}
	\author{Sourav Dutta$^2$\footnote {sduttaju@gmail.com}}
	\author{Subenoy Chakraborty$^1$\footnote {schakraborty.math@gmail.com}}
	\affiliation{$^1$Department of Mathematics, Jadavpur University, Kolkata-700032, West Bengal, India\\$^2$Department of Mathematics, Dr. Meghnad Saha College, Itahar, Uttar Dinajpur-733128, West Bengal, India.}

	
	\begin{abstract}
		In the framework of $f(T)$-gravity theory, classical and quantum cosmology has been studied in the present work for FLRW space-time model. The Noether symmetry, a point-like symmetry of the Lagrangian is used to the physical system and a specific functional form of $f(T)$ is determined. A point transformation in the 2D augmented space restricts one of the variable to be cyclic so that the Lagrangian as well as the field equations are simplified so that they are solvable. Lastly for quantum cosmology, the WD equation is constructed and possible solution has been evaluated.	
	\end{abstract}
	\maketitle

	\section{Introduction}
In the context of recent series of observational evidences \cite{Riess:1998cb, Perlmutter:1998np, Spergel:2003cb, Tegmark:2003ud, Eisenstein:2005su} which predict that our Universe is going through an era of accelerated expansion, a group of cosmologists are in favour of modifying Einstein gravity to accommodate these predictions. There are several modified gravity theories in the literature among which the popular one is the $f(R)$-gravity theory \cite{Capozziello:2003tk, Carroll:2003wy}. In this gravity theory, the scalar curvature $R$ in the Einstein-Hilbert action is replaced by an arbitrary function $f(R)$. In recent years, another gravity theory gets much attention and is known as teleparallel gravity. Here the gavitational interactions \cite{Einstein/00, Einstein/01, Hayashi/00} are described by torsion (instead of curvature). Such a gravity model was first proposed by Einstein with a view to unify electromagnetism and gravity over Weitzenb$\ddot{\mbox{o}}$ck non-Riemannian manifold. So the Levi-civita connection is replaced by Weitzenb$\ddot{o}$ck connection in the underlying Riemann-cartan space-time. As a result, pure geometric nature of the gravitational interaction is violated and torsion behaves as force. Hence gravity may be considered as a gauge theory of the translation group \cite{Arcos:2005ec}. Although there are conceptual differences between GR and teleparallel gravity theory, still at classical level both of them have equivalent dynamics.

In analogy to $f(R)$-gravity theory, a generalization to teleparallel gravity has been formulated \cite{Ferraro:2006jd, Ferraro:2011ks, Linder:2010py, Finch/00, capozziello/00, Basilakos/00, Bamba:2013jqa, Li:2018ixg, Basilakos/01, Harko:2014sja} by replacing the torsion scalar $T$ by a generic function $f(T)$. Linder \cite{Linder:2010py} termed this modified gravity as $f(T)$-gravity theory. For a comparative study with $f(R)$-gravity theory, there are two important differences namely (a) the field equations in $f(T)$-gravity theory are second order while one has fourth order equations in $f(R)$-gravity \cite{Bengochea:2008gz} (b) Although $f(R)$-gravity theory obey local lorentz invariance but not by $f(T)$-gravity theory. As a result, in $f(T)$-gravity theory all $16$ components of the vierbien are independent and a gauge choice \cite{Li:2010cg} can not fix six of them. Further, the four linearly independent vierbeins (i.e, tetrad) fields are the dynamical object in $f(T)$-gravity theory. Also these tetrad fields form the orthogonal bases for the tangent space at each point of space-time. The name ``teleparallel" is justified as the vierbeins are parallel vector fields (for a review see ref. \cite{Cai:2015emx, Darabi:2019qpz, Golovnev/00, Aviles/00, Bamba/02, Mishra/00, Biswas/00}).

The geometrical symmetries namely Lie point and Noether symmetries related to space-time are usually very useful to solve/study physical systems. The conserved charges in Noether symmetry  are considered as a selection criterion to discriminate similar physical processes \cite{Capozziello/100000, Szydlowski/10000000, Capozziello:2007wc, Capozziello/50000000, Capozziello:2008ch, Capozziello:2009te, Vakili:2008ea, Zang/10000}. Mathematically a given system of differential equation can either be simplified or to have a first integral (Norther integral) by imposing Noether symmetry to the system. Further, it is possible to constrain or determine physical parameters involved in a physical system by imposing Noether symmetry to it \cite{Szydlowski/00, Dutta:2016jgj}. In recent years, symmetry analysis has been widely used to the physical systems in Riemannian spaces \cite{Tsamparlis/11, Leach/11, Bluman/11, Aminova/11, Aminova/12, Feroze/11, Tsamparlis/111} and specially in the context of cosmology \cite{Dutta/00, Dutta/01, Dutta:2016exd, Dutta:2016mbp, Dutta:2018hyv, Dutta:100}. The present work is an example of it. Usually, evolution equations are simplified to a great extend by determining a cyclic variable in the augmented space. As a consequence, analytic solutions are possible with new variables (in the augmented space) and are analyzed from cosmological context.

On the otherhand, Noether symmetry can also be used in quantum cosmology to identify a typical subset of the general solution of the Wheeler-Dewitt (WD) equation having oscillatory behavior \cite{Dutta:101, Dutta:2019ujc, Paliathanasis:2015gga, Capozziello:2012hm,Basilakos:2013rua,Bajardi:2021tul}. Also in Minisuperspace geometry, symmetry analysis identifies equations of classical trajectories \cite{Capozziello:2013qha, Halliwell/00}. Hence classical observable universe can be related to the quantum cosmology through the application of Noether symmetry analysis.

In the present work, both classical and quantum cosmology has been studied for $f(T)$-gravity theory using the Noether symmetry analysis in the background of homogeneous and isotropic flat FLRW space-time model. The paper is organized as follows: a basic concept of Noether symmetry approach has been discussed in Section II, Section III describes the brief review of conformal symmetry and classical cosmology in f(T) gravity and Noether symmetry is presented in Section IV, where as Section V presents a general description of quantum cosmology: minisuperspace approach and paper ends with a summary in Section VI.
\section{Basic Concept of Noether Symmetry approach}
Noether's first theorem states that every differentiable symmetry of the action of a physical system with conservative forces has a corresponding conservation law i.e, the lie derivative of the Lagrangian \cite{Paliathanasis:2015gga}, \cite{Basilakos:2011rx}, \cite{Paliathanasis:2014zxa}  of any physical system, associated to some conserved quantities, will be invariant along an appropriate vector field ($\mathcal{L}_{\overrightarrow{X}}f=\overrightarrow{X}(f)$). Using these one can determine the conserved quantities (invariants) from the observed symmetries of a physical system. Also these symmetry constraints  are very much useful to simplify the evolution equations of the physical system \cite{Stephani/00}.

If a point like canonical Lagrangian is of the form $L[q^{\alpha}(x^j),\dot{q}^{\alpha}(x^j)]$, then the Euler-Lagrange equations take the form 
\begin{equation}
	\partial_{i}\left(\frac{\partial L}{\partial\partial_{i}q^{\alpha}}\right)=\frac{\partial L}{\partial q^{\alpha}}\label{eq1}
\end{equation}
Now contracting the equation (\ref{eq1}) with some unknown functions $\lambda^{\alpha}(q^{\beta})$ one can get 
\begin{equation}
	\lambda^{\alpha}\bigg[\partial_{i}\left(\frac{\partial L}{\partial\partial_{i}q^{\alpha}}\right)-\frac{\partial L}{\partial q^{\alpha}}\bigg]=0\label{eq2}
\end{equation}
\begin{equation}
	\lambda^{\alpha}\frac{\partial L}{\partial q^{\alpha}}+(\partial_{i}\lambda^{\alpha})\left(\frac{\partial L}{\partial\partial_{i}q^{\alpha}}\right)=\partial_{i}\left(\lambda^{\alpha}\frac{\partial L}{\partial\partial_{i}q^{\alpha}}\right)\nonumber
\end{equation}
So, the Lie derivative of the Lagrangian i.e, $\mathcal{L}_{\overrightarrow{X}}L$ takes the form,
\begin{equation}
	\mathcal{L}_{\overrightarrow{X}}L=\lambda^{\alpha}\frac{\partial L}{\partial q^{\alpha}}+(\partial_{i}\lambda^{\alpha})\frac{\partial L}{\partial\left(\partial_{i}q^{\alpha}\right)}=\partial_{i}\left(\lambda^{\alpha}\frac{\partial L}{\partial\partial_{i}q^{\alpha}}\right)\label{eq3}
\end{equation}
where $\overrightarrow{X}$, the infinitesimal generator, is of the form 
\begin{equation}
	\overrightarrow{X}=\lambda^{\alpha}\frac{\partial}{\partial q^{\alpha}}+\left(\partial_{i}q^{\alpha}\right)\frac{\partial}{\partial\left(\partial_{i}q^{\alpha}\right)}\label{eq4}
\end{equation}
According to Noether theorem, the lie derivative of the Lagrangian $L$ along the vector field $\overrightarrow{X}$ must vanish i.e, $\mathcal{L}_{\overrightarrow{X}}L=0$.
Further,  from equation (\ref{eq3}) we can say that the Noether current or conserved current $Q^i$ is a constant of motion of the system \cite{Capozziello:2010ih}, where 
\begin{equation}
	Q^i=\lambda^{\alpha}\frac{\partial L}{\partial\left(\partial_{i}q^{\alpha}\right)}\label{eq5}
\end{equation}
satisfying
\begin{equation}
	\partial_{i}Q^i=0\label{eq6}
\end{equation}
Now the energy function associated with the system can be written as 
\begin{equation}
	E=\dot{q}^{\alpha}\frac{\partial L}{\partial\dot{q}^{\alpha}}-L\label{eq7}
\end{equation}
If there is no explicit time dependence in the Lagrangian, then the energy function which is also known as Hamiltonian of the system is a constant of motion \cite{Capozziello:2010ih}. In section 4, we will show how Noether symmetry analysis simplifies the present coupled cosmological model. 

Hamiltonian formulation is very useful in the context of quantum cosmology. The Noether symmetry condition can be rewritten as 
\begin{equation}
	\mathcal{L}_{\overrightarrow{X}_{H}}H=0\label{eq8}
\end{equation}
with
\begin{equation}
	{\overrightarrow{X}_{H}}=\dot{q}\frac{\partial}{\partial q}+\ddot{q}\frac{\partial}{\partial\dot{q}}\nonumber
\end{equation}
The conserved canonically conjugate momenta due to Noether symmetry can be written as 
\begin{equation}
	\Pi_{l}=\frac{\partial L}{\partial q^l}={\sum}_{l}\label{eq9}
\end{equation}
$l=1,2,...m$

where '$m$' is the no of symmetries. Now the operator version (i.e, quantization) of equation (\ref{eq9}) takes the form
\begin{equation}
	-i\partial_{q^l}\ket{\psi}={\sum}_{l}\ket{\psi}\label{eq10}
\end{equation}
For real conserved quantity $\sum_{l}$, the equation (\ref{eq10}) has oscillatory solutionwhich is given by
\begin{equation}
	\ket{\psi}=\sum_{l=1}^{m}e^{i\sum_{l}q^l}\ket{\phi(q^k)}  , k<n\label{eq11}
\end{equation}
where '$k$' is the directions along which there is no symmetry and $n$ is the dimension of the minisuperspace. This oscillatory part of the wave function implies that the Noether symmetry exists and the conjugate momenta along the symmetry directions should be conserved and vice-versa (Hartle \cite{Hartle/1000}). In fact Noether symmetry allows researchers to consider whole classes of hypothetical Lagrangians with given invariants to describe a physical system.
\section{Conformal Symmetry: A brief review}
In differential geometry, conformal invariance gives rich geometrical structures. A vector field $\xi^{\alpha}$ is a Conformal Killing Vector (CKV) of the metric $g_{ij}$ if 
\begin{equation}
	\mathcal{L}_{\overrightarrow{\xi}}g_{ij}=\mu(x^k)g_{ij}\label{eq12}
\end{equation}
where $\mu$ is an arbitrary function of the space and notationally $\mathcal{L}_{\overrightarrow{\xi}}$ indicates Lie derivative with respect to the vector field $\overrightarrow{\xi}$. In particular if 

(i)~$\mu(x^k)=\mu_{0}(\neq 0)$, a constant: $\xi^{\alpha}$ - homothetic vector field.

(ii)~$\mu(x^k)=0$: $\xi^{\alpha}$ - killing vector field.

The above three class of vector fields individually form an algebra as follows:

(a)~The class of conformal killing vectors form an algebra, known as conformal algebra (CA) of the metric \cite{Tsamparlis:2013aza}.

(b)~The class of homothetic vector fields form an algebra, known as homothetic algebra (HA).

(c)~The class of killing vector fields form an algebra, known as killing algebra (KA).

These three algebras are related as 
\begin{equation}
	KA\subseteq HA\subseteq CA\label{eq13}
\end{equation}
Further, for a $n(>2)$ dimensional manifold of constant curvature, the dimension of these three algebras are $\dfrac{(n+1)(n+2)}{2}$, $\dfrac{n(n+1)}{2}+1$ and $\dfrac{n(n+1)}{2}$ respectively.

In a given space, two metrics $g$ and $g'$ are said to be conformally related if $\exists$ a function $\Pi(x^k)$ so that 
\begin{equation}
	g'_{ij}=\Pi(x^k)g_{ij}\label{eq14}
\end{equation}
It is to be noted that two conformally related metrics have the same conformal algebra but subalgebras are not necessarily same. In fact if $\overrightarrow{\xi_{0}}$ is a conformal killing vector for the conformally related metrics $g$ and $g'$ then the corresponding conformal functions $\mu(x^k)$ and $\mu'(x^k)$ are related by the relation 
\begin{equation}
	\mu'(x^k)=\mu(x^k)+\mathcal{L}_{\overrightarrow{\xi_{0}}}(\ln\Pi)\label{eq15}
\end{equation}
As Noether symmetries follow the homothetic algebra of the metric so two conformally related physical systems are not identical.

In the context of conformal Lagrangian, Tsamparlis et al. \cite{Tsamparlis:2013aza} have shown that the equations of motion (i.e, Euler-Lagrange equations) corresponding to two conformal Lagrangians transform covariantly under the conformal transformation provided Hamiltonian (i.e, the total energy) is zero. So systems with vanishing energy are conformally related and corresponding equations of motion are conformally invariant. Further, in quantum cosmology, due to the the Hamiltonian constraint the total energy of the system has be zero, and consequently, one has conformally invariant systems with respect to the equations of motion.
\section{Classical Cosmology in $f(T)$-gravity and  Noether Symmetry}
In the background of flat FLRW space-time model, the point like Lagrangian in $f(T)$ gravity theory takes the form 
\begin{equation}
	L=a^3f(T)-a^3Tf_{T}(T)-6a\dot{a}^2f_{T}(T)-Da^{-3\omega}\label{eq16}
\end{equation}  
where $f(T)$ is a regular function of the torsion scalar $T$, $a$ is the scale factor and $D$ is a constant of integration. Here matter field is chosen as perfect fluid with $\omega=\dfrac{p}{\rho}$, the constant equation of state parameter. The modified Friedmann equations are \cite{Bose:2020xdz}
\begin{equation}
	H^2=\frac{1}{(2f_{T}+1)}\left[\frac{\rho}{3}-\frac{f}{6}\right]\label{eq17}
\end{equation} 
and
\begin{equation}
	2\dot{H}=\frac{(\rho+p)}{1+f_{T}+2Tf_{TT}}\label{eq18}
\end{equation}
In the present Lagrangian system we have 2D configuration space $\{a,T\}$ and the momenta conjugate to configuration variables are
\begin{eqnarray}
	p_{a}&=&\frac{\partial L}{\partial\dot{a}}=-12a\dot{a}f_{T}(T)\label{eq19}\\
	p_{T}&=&\frac{\partial L}{\partial\dot{T}}=0\label{eq20}
\end{eqnarray}
Using Legendre transformation, the Hamiltonian of the system is 
\begin{equation}
	H=-\frac{1}{24}\frac{{p_{a}}^2}{af_{T}(T)}-a^3f(T)+a^3Tf_{T}(T)+Da^{-3\omega}\label{eq21}
\end{equation}
So the Hamilton's equations of motion are
\begin{eqnarray}
	\dot{a}&=&\{a,H\}=-\frac{1}{12}\frac{p_{a}}{af_{T}(T)}\nonumber\\
	\dot{T}&=&\{T,H\}=0\nonumber\\
	\dot{p_{a}}&=&\{p_{a},H\}=-\frac{1}{24}\frac{{p_{a}}^2}{a^2f_{T}(T)}+3a^2f(T)-3a^2Tf_{T}(T)+3\omega Da^{-3\omega-1}\nonumber\\
	\dot{p_{T}}&=&\{p_{T},H\}=6a\dot{a}^2f_{TT}(T)-a^3Tf_{TT}(T)\label{eq22}
\end{eqnarray}
We shall now impose Noether symmetry to the above physical system. According to Noether's theorem $\exists$ a vector field $\overrightarrow{X}$ about which the Lie derivative of the Lagrangian should be zero i.e, 
\begin{equation}
	\mathcal{L}_{\overrightarrow{X}}L=0\label{eq23}
\end{equation}
where the infinitesimal generator $\overrightarrow{X}$ has the form 
\begin{equation}
	\overrightarrow{X}=\alpha\frac{\partial}{\partial a}+\beta\frac{\partial}{\partial T}+\dot{\alpha}\frac{\partial}{\partial \dot{a}}+\dot{\beta}\frac{\partial}{\partial\dot{T}}\label{eq24}
\end{equation}
Here $\alpha=\alpha(a,T)$ and $\beta=\beta(a,T)$ are the functions in the configuration space with $\dot{\alpha}=\dfrac{\partial\alpha}{\partial a}\dot{a}+\dfrac{\partial\alpha}{\partial T}\dot{T}$ and so on.

Now from the Noether symmetry condition (\ref{eq23}) one obtains the following partial differential equations:
\begin{equation}
	-6\alpha f'(T)-6a\beta f''(T)-12af'(T)\frac{\partial\alpha}{\partial a}=0\label{eq25}
\end{equation}
\begin{equation}
	-12af'(T)\frac{\partial\alpha}{\partial T}=0\label{eq26}
\end{equation}
and
\begin{equation}
	3\alpha a^2f(T)-3\alpha a^2Tf'(T)+3\omega D\alpha a^{-3\omega-1}-\beta a^3f''(T)T=0\label{eq27}
\end{equation}
Using seperation of variables for the coefficients $\alpha$, $\beta$ of the symmetry vector, the above set of partial differential equations are solvable to give 
\begin{eqnarray}
	\alpha(a,T)&=&ca^{1-\frac{3k}{2}}\nonumber\\
	\beta(a,T)&=&-3kca^{-\frac{3k}{2}}T\label{eq28}
\end{eqnarray}
and also $f(T)$ has the solution 
\begin{equation}
	f(T)=f_{0}T^{\frac{1}{k}}\label{eq29}
\end{equation}
with the equation of state parameter $\omega=0$.

In the solution, $c$, $k$ and $f_{0}$ are arbitrary integration constants.

In order to solve the modified Friedmann equations we make a transformation in the configuration space $(a,T)\rightarrow(u,v)$ so that one of the transformed variable (say $u$) becomes cyclic and consequently the transformed Lagrangian becomes much simpler in form. Hence the evolution equations become very simple to have analytic solutions. So the infinitesimal generator (i.e, the vector field $\overrightarrow{X}$) due to this point transformation becomes 
\begin{equation}
	\overrightarrow{X}_{T}=\left(i_{\overrightarrow{X}}du\right)\frac{\partial}{\partial u}+\left(i_{\overrightarrow{X}}dv\right)\frac{\partial}{\partial v}+\left\{\frac{d}{dt}\left(i_{\overrightarrow{X}}du\right)\right\}\frac{\partial}{\partial\dot{u}}+\left\{\frac{d}{dt}\left(i_{\overrightarrow{X}}dv\right)\right\}\frac{\partial}{\partial\dot{v}}\label{eq30}
\end{equation}
Thus $\overrightarrow{X}_{T}$ may be considered as the lift of a vector field defined on the augmented space. Now, without any loss of generality one may restrict the above point transformation to be
\begin{equation}
	i_{\overrightarrow{X}}du=1\nonumber
\end{equation}	
and
\begin{equation}
	i_{\overrightarrow{X}}dv=0\label{eq31}
\end{equation}
so that
\begin{equation}
	\overrightarrow{X}_{T}=\dfrac{\partial}{\partial u}\nonumber
\end{equation}  
with
\begin{equation}
	\dfrac{\partial L_{T}}{\partial u}=0\label{eq32}
\end{equation}
Here $i_{\overrightarrow{X}}$ stands for the inner product with the vector field $\overrightarrow{X}$. Usually with Noether symmetry there is an associated conserved current (defined in the (\ref{eq5}) of section 2). The time component of it when integrated over spatial volume gives a conserved charge, which in geometric notion can be defined as 
\begin{equation}
	q=i_{\overrightarrow{X}}\theta_{L}\nonumber
\end{equation}
where the cartan one form $\theta_{L}$ is defined as \cite{Dutta:2018hyv} 
\begin{equation}
	\theta_{L}=\frac{\partial L}{\partial a}da+\frac{\partial L}{\partial T}dT\label{eq33}
\end{equation}
Now the first order partial differential equations corresponding to equation (\ref{eq31}) are
\begin{equation}
	\alpha\frac{\partial u}{\partial a}+\beta\frac{\partial u}{\partial T}=1\nonumber
\end{equation}
and
\begin{equation}
	\alpha\frac{\partial v}{\partial a}+\beta\frac{\partial v}{\partial T}=0\label{eq34}
\end{equation}
Using the solutions for $\alpha$ and $\beta$ from equation (\ref{eq28}) the solution for $u$ and $v$ gives
\begin{equation}
	u=\frac{2}{3kc}a^{\frac{3k}{2}}\nonumber
\end{equation}
\begin{equation}
	v=\ln\left(aT^{\frac{1}{3k}}\right)\label{eq35}
\end{equation}
and the transformed Lagrangian takes the form 
\begin{equation}
	L=f_{0}\left(1-\frac{1}{k}\right)e^{3v}-\frac{6f_{0}}{k}c^2\dot{u}^2e^{\left(3-3k\right)v}-D\label{eq36}
\end{equation}
The solution of the corresponding Euler-Lagrange equations take the form 
\begin{eqnarray}
	v&=&B\nonumber\\
	u&=&Ft+G\label{eq37}
\end{eqnarray}
with $B$, $F$ and $G$ are arbitrary constants.
\begin{figure}
	\includegraphics[height=.5\textheight,width=0.6\textheight]{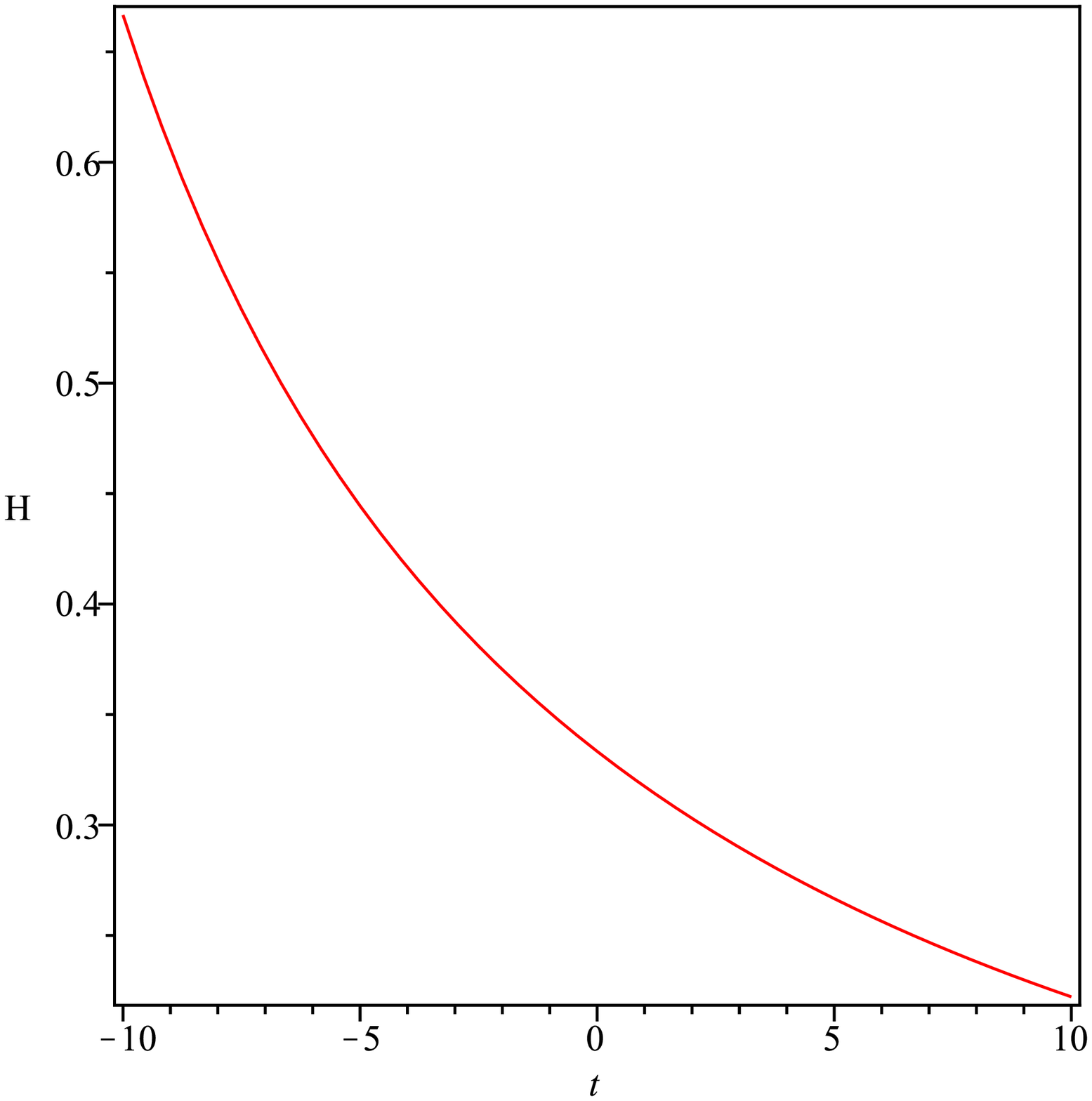}
	\caption{$H$ vs $t$ }  
\end{figure}\label{f1}
\newpage
\begin{figure}
	\includegraphics[height=.5\textheight,width=0.6\textheight]{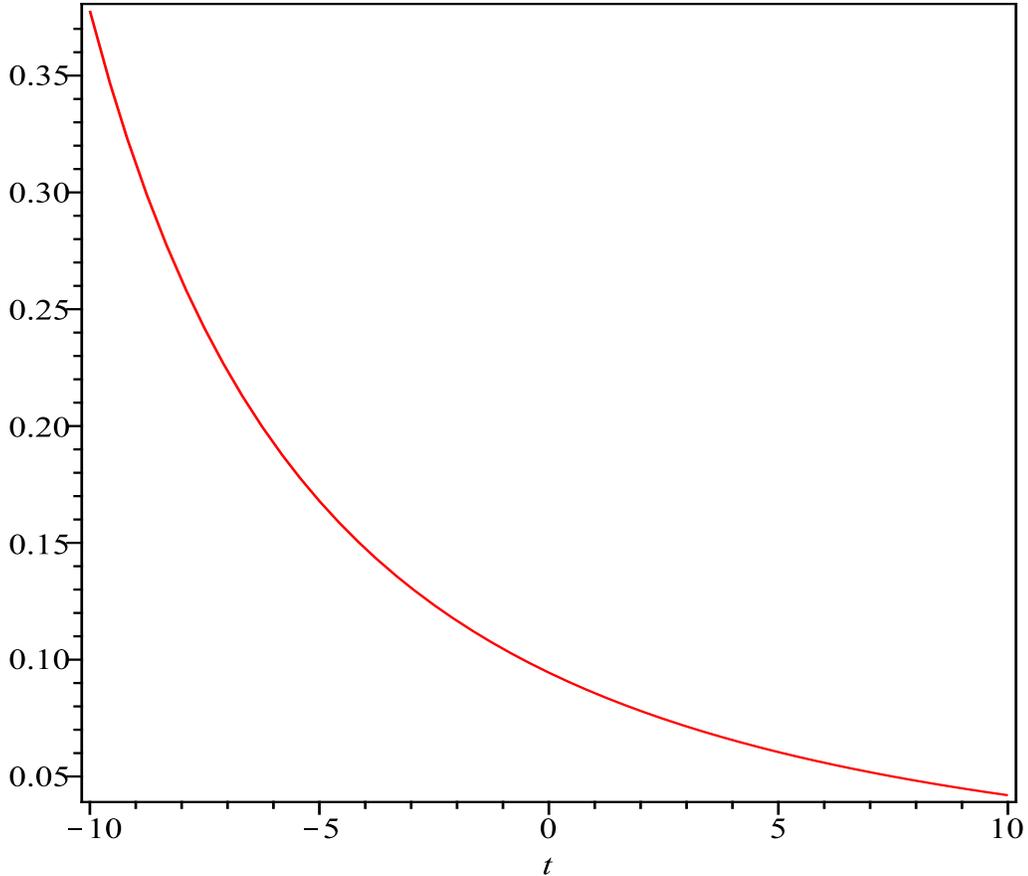}
	\caption{$\frac{\ddot{a}}{a}$ vs $t$}  
\end{figure}\label{f2}
So the classical cosmological solution in original variables can be written as 
\begin{eqnarray}
	a&=&\left\{\frac{3kc}{2}\left(Ft+G\right)\right\}^{\dfrac{2}{3k}}\nonumber\\
	T&=&\frac{4B_{0}}{9k^2c^2\left(Ft+G\right)^2}\nonumber\\
	f(T)&=&f_{0}\left\{\frac{4B_{0}}{9k^2c^2\left(Ft+G\right)^2}\right\}^{\dfrac{1}{k}}\label{eq38}
\end{eqnarray}
where $B_{0}$ is an arbitrary constant.

The above cosmological solution indicates power-law expansion of the Universe with Hubble parameter decreases with respect to cosmic time as $\dfrac{1}{t}$ (see Figure 1) and the Universe is in an accelerating phase (see Figure 2) with rate of acceleration decreases with the evolution.

\section{A general description of Quantum Cosmology: Minisuperspace Approach}

In cosmology, homogeneous and isotropic metrics and matter fields are the simplest and widely used minisuperspace models. In this model lapse function $N(=N(t))$ is homogeneous and the shift function identically vanishes. Using $(3+1)$-decomposition the metric in $4D$ manifold takes the form,
\begin{equation}
	ds^2=-N^2(t)dt^2+h_{ab}(x,t)dx^adx^b\label{eq39}
\end{equation}   
and the Einstein-Hilbert action can be written as 
\begin{equation}
	I(h_{ab},N)=\frac{{m^2}_p}{16\pi}\int dtd^3xN\sqrt{h}\left[k_{ab}k^{ab}-k^2+(3)_{R}-2\Lambda\right]\label{eq40}
\end{equation}
here $k_{ab}$ denotes the extrinsic curvature of the $3$ space; the trace of the extrinsic curvature $k=k_{ab}h^{ab}$; $(3)_{R}$ represents the curvature scalar of the three space and $\Lambda$ is a constant known as cosmological constant.

The metric $h_{ab}$ can be characterized by a finite number of time functions $q^{\alpha}(t)$, $\alpha=0,1,2,...,n-1$ due to homogeneity of the three space. So, the above action takes the form of a relativistic point particle with self interacting potential in $nD$ curved space time as
\begin{equation}
	I\left(q^{\alpha}(t),N(t)\right)=\int_{0}^{1}dtN\left[\frac{1}{2N^2}f_{\alpha\beta}(q)\dot{q}^{\alpha}\dot{q}^{\beta}-V(q)\right]\label{eq41}
\end{equation} 
So the equation of motion of the relativistic particle can be written as 
\begin{equation}
	\frac{1}{N}\frac{d}{dt}\left(\frac{\dot{q}^{\alpha}}{N}\right)+\frac{1}{N^2}\Gamma^{\alpha}_{\mu\nu}\dot{q}^{\mu}\dot{q}^{\nu}+f^{\alpha\beta}\frac{\partial\nu}{\partial q^{\beta}}=0\label{eq42}
\end{equation}
where $\Gamma^{\alpha}_{\beta\gamma}$ is the Christoffel symbols in the minisuperspace. Now we also have a constraint equation which is obtained by variation with respect to the lapse function as 
\begin{equation}
	\frac{1}{2N^2}f_{\alpha\beta}\dot{q}^{\alpha}\dot{q}^{\beta}+V(q)=0\label{eq43}
\end{equation}
Now the momenta canonical to $q^{\alpha}$ can be written as 
\begin{equation}
	p_{\alpha}=\frac{\partial L}{\partial q^{\alpha}}=f_{\alpha\beta}\frac{\dot{q}^{\beta}}{N},\label{eq44}
\end{equation}
so the Hamiltonian can be written as
\begin{equation}
	H=p_{\alpha}\dot{q}^{\alpha}-L=N\left[\frac{1}{2}f^{\alpha\beta}p_{\alpha}p_{\beta}+V(q)\right]=N\mathcal{H}\label{eq45}
\end{equation}
where $f^{\alpha\beta}$ is the inverse metric. Using equation (\ref{eq43}) and equation (\ref{eq44}) we get
\begin{equation}
	\mathcal{H}(q^{\alpha},p_{\alpha})\equiv\frac{1}{2}f^{\alpha\beta}p_{\alpha}p_{\beta}+V(q)=0\label{eq46}
\end{equation}
Now, in quantization scheme, replacing $p_{\alpha}$ by $-i\hbar\dfrac{\partial}{\partial q_{\alpha}}$ the operator version of equation (\ref{eq46}) on a time independent function one can obtain the Wheeler-Dewitt (WD) equation in quantum cosmology as 
\begin{equation}
	\mathcal{H}\left(q^{\alpha},-i\hbar\frac{\partial}{\partial q^{\alpha}}\right)\psi(q^{\alpha})=0\label{eq47}
\end{equation}
The above WD equation encounters the operator ordering problem because generally the minisuperspace metric is dependent on $q^{\alpha}$. One can resolve this problem by imposing the quantization in minisuperspace to be covariant in nature. In the context of quantum cosmology for probability measure, $\exists$ a conserved current for hyperbolic type of partial differential equation as  
\begin{equation}
	\overrightarrow{J}=\frac{i}{2}(\psi^{*}\nabla\psi-\psi\nabla\psi^{*})\label{eq48}
\end{equation}
and $\overrightarrow{J}$ satisfies the relation $\overrightarrow{\nabla}.\overrightarrow{J}=0$. Here $\psi$ represents the solution of WD differential equation (hyperbolic type). So the probability measure on the minisuperspace can be written as 
\begin{equation}
	dp=|\psi(q^{\alpha})|^2dV\label{eq49}
\end{equation}
where $dV$ denotes a volume element on minisuperspace.

In the present problem the Lagrangian of the system in the transformed variables is given by (\ref{eq36}). So the canonically conjugate momenta are 
\begin{eqnarray}
	p_{u}&=&\frac{\partial L}{\partial\dot{u}}=-\frac{12f_{0}}{k}c^2\dot{u}e^{(3-3k)v}\nonumber\\
	p_{v}&=&\frac{\partial L}{\partial\dot{v}}=0\label{eq50}
\end{eqnarray}
Hence the Hamiltonian of the system is 
\begin{equation}
	H=p_{u}\dot{u}+p_{v}\dot{v}-L=-\frac{k}{24f_{0}c^2}e^{(3k-3)v}p_{u}^2-f_{0}\left(1-\frac{1}{k}\right)e^{3v}+D\label{eq51}
\end{equation}

The above Hamiltonian is a very special type of Hamiltonian having only one dynamical variable $u$ which is also cyclic in nature ($v$ can not be considered as dynamical variable as $\dot{v}$ i.e, $p_{v}$ does not appear in the Hamiltonian). 

Hence the WD equation which is the operator version of the above Hamiltonian takes the form 
\begin{equation}
	\frac{d^2\phi}{du^2}+l\phi=0\label{eq52}
\end{equation}
with $l=\dfrac{f_{0}\left(1-\frac{1}{k}\right)e^{3v}-D}{\frac{k}{24f_{0}c^2}e^{(3k-3)v}}$, a constant.\\

So the solution of WD equation can be written as 
\begin{eqnarray}
	\phi&=&A_{1}e^{\sqrt{l}u}+A_{2}e^{-\sqrt{l}u},~~ \mbox{when}~ l>0\nonumber\\
	&=&B_{1}\cos\sqrt{-l}u+B_{2}\sin\sqrt{-l}u,~~\mbox{when}~ l<0\nonumber\\
	&=&C_{1}u+C_{2},~~\mbox{when} ~l=0\label{eq55}
\end{eqnarray}
where $A_{i}$'s, $B_{i}$'s and $C_{i}$'s $(i=1,2)$ are the constants of integration.

\section{Summary}
The present work deals with $f(T)$ cosmology from the point of view of symmetry analysis. In particular Noether symmetry has been used both in classical and quantum cosmology with $f(T)$ gravity theory. Using Noether symmetry condition to the Lagrangian of the present model along the symmetry vector, the coefficients of the symmetry vector are not only determined, it is possible to determine the explicit form of the $f(T)$ function. Using a transformation of variables in the configuration (satisfying condition $(\ref{eq31})$) the Lagrangian simplifies to a great extend and it is possible to have cosmological solution having power-law nature. In quantum cosmology, the WD equation simplifies to a great extend due to only one dynamical variable having cyclic nature. From the nature of the wave function one can infer that big-bang singularity may be avoided quantum mechanically for the present $f(T)$-cosmological model. So one may conclude that $f(T)$ cosmology can avoid the big-bang singularity at the very early era of evolution of the Universe.


\end{document}